\documentclass{article}

\usepackage[dvips]{graphicx}

\usepackage[utf8]{inputenc}
\usepackage{fancybox} 
\usepackage{tabularx} 
\usepackage{multirow}
\usepackage{multicol}
\usepackage{booktabs}
\usepackage{grffile}
\usepackage{xcolor}

\newcommand{\nn}{\nonumber}
\newcommand{\be}{\begin{equation}}
\newcommand{\ee}{\end{equation}}
\newcommand{\ba}{\begin{eqnarray}}
\newcommand{\ea}{\end{eqnarray}}
\newcommand{\req}[1]{(\ref{#1})}
\def\={\,=\,}
\newcommand{\ci}[1]{\cite{#1}}

\def\mev{~{\rm MeV}}
\def\gev{~{\rm GeV}}

\def\LQCD{\Lambda_{\rm QCD}}

\newcommand{\tw}{\textwidth}

\def\vb0{{\bf b}_0}

\def\={\,=\,}

\begin{document} 
\thispagestyle{empty}
\begin{flushright}
December, 10 2019 \\[20mm]
\end{flushright}

\begin{center}
  {\Large\bf The role of the GPD $E$ in the determination of the parton angular momenta}\\
            \vskip 15mm
P.\ Kroll \\[1em]
{\small {\it Fachbereich Physik, Universit\"at Wuppertal, D-42097 Wuppertal
Germany }}\\[1em]
\end{center}

\begin{abstract}
For the generalized parton distribution (GPD) $E$, extracted from hard exclusive processes
and the Pauli form factor of the nucleon, the second moments for quarks and gluons at $t=0$
are calculated and their scale dependence analyzed. The parton angular momenta are subsequently
determined and compared with other results.
\end{abstract}
\section{Introduction}
One of the goals of investigating hard exclusive processes is the determination of the 
parton angular momenta. According to Ji \ci{ji96} the (total) angular momentum, $J$, which 
partons inside a proton carry, is given by half of the sum of the second moments of the GPDs 
$H$ and $E$ at zero invariant momentum transfer, $t$, and zero skewness, $\xi$, the ratio of 
the difference and the sum of the light-cone plus components of the outgoing and ingoing proton 
momenta. Thus, for quarks of flavor $a$ and their antiquarks, the angular momentum is
\be
J^{a+\bar{a}}\=\frac12\,\int_{-1}^1 dx x\big[H_a(x,\xi=t=0)+E_a(x,\xi=t=0)\big]\,.
\label{eq:J-def}
\ee
An analogous relation holds for the gluon. 
Strictly speaking, for a proton that moves along the 3-direction, $J$ is the expectation value of 
the 3-component of the parton angular momentum operator. For detailed discussions of the 
definition and interpretation of the parton angular momenta see the reviews \ci{diehl03} and 
\ci{leader13}. The parton angular momenta have found tremendous interest in connection with the 
so-called 'spin crisis' which has its origin in comparing quark spin information at low scales 
with results from deep inelastic lepton-nucleon scattering (DIS) obtained at much larger scales. 
As became recently clear the scale dependence, i.e.\ the evolution of the relevant quantities 
with the scale, explains the 'spin crisis' \ci{thomas,haegler10}.

While the forward limit of the GPD $H$ reduces to the usual parton density (PDF), known from
DIS, the GPD $E$ is the genuine contribution of hard exclusive processes to the issue of the 
angular momenta. The present paper is addressed to the extraction of $E$ from data on such processes.
For valence quarks $E$ has already been determined from the
data on the Pauli form factor of the nucleon \ci{DFJK4,DK13} and is used in several calculations of the
transverse spin asymmetry, $A_{UT}$, in various hard exclusive processes \ci{GK4,GK7,KMS}. Here, in this
article use will be made of the results derived in \ci{DK13}. As will be shown below, using a positivity
bound on $E$ for strange quarks, $E^s$, derived in \ci{diehl-kugler}, together with DVCS data on
$A_{UT}$ \ci{hermes-aut-dvcs}, information on $E^s$ can be extracted although with large uncertainties.
For the extracted GPD $E$ the second moments are subsequently evaluated and their evolution properties
studied. Finally, these moments of $E$ in combination with the second moments of the parton density,
taken from \ci{ABM11}, allow for a computation of the parton angular momenta in dependence
of the scale.
  
The layout of this paper is as follows: In Sect.\ 2 the definitions of moments of the GPD
are presented and the relevant sum rules are recapitulated. In Sect.\ 3 the present knowledge
of the GPD $E$ is examined and in Sect.\ 4, it is dealt with the evolution  of its 2nd moments.
Sect.\ 5 is addressed to the determination of the parton angular momenta. Finally, in Sect.\ 6, the
summary is presented followed by a brief discussion of implications of the results on $J$
for the orbital angular momenta carried the partons inside the proton.

\section{The sum rules}
\label{sec:sum-rules}

The second Mellin moments of the densities for quarks and gluons which are equivalent to
the forward limits of the corresponding GPD $H$, are defined as  
\be
q_{20}^a\=\int_0^1 dx x q^a(x)\,, \qquad g_{20}\=\int_0^1dx x g(x)\,,
\label{eq:density-moments}
\ee
where here and in the following $a$ denotes one of the quark flavors
\be
a\= u, d, s, \bar{u}, \bar{d}, \bar{s}, \ldots
\ee
The second moments of the GPD $E$ at $\xi=t=0$ are defined by~\footnote{
      The $n$-th moment of a GPD $K(x,\xi,t)$ is an even polynomial in $\xi$ of
      order $n$. The coefficients, generalized form factors, are denoted by  $k_{ni}(t)$ \ci{diehl03}.
      $k_{n0}$ is the $\xi$ independent term of that polynomial.}
\be
e_{20}^a\=\int_0^1dx x E^a(x,\xi=t=0)\,, \qquad e_{20}^g\=\int_0^1 dx E^g(x,\xi=t=0)\,.
\label{eq:e-moments}
\ee
All moments and therefore the angular momenta too, depend on the renormalization scale, $\mu$. 
The reader is reminded that the gluon GPD $E^g$ is defined in analogy to the gluon GPD, $H^g$, for
  which the $\xi, t\to 0$ limit is $xg(x)$, see for instance \ci{diehl03}. 

With these definitions the angular momenta can be expressed as
\be
J^a\=\frac12\big[q_{20}^a + e_{20}^a\big]\,, \qquad J^g\=\frac12[g_{20} + e_{20}^g\big]\,.
\label{eq:ji}
\ee 
The sum of all parton angular momenta is $1/2$, the spin of the proton:
\be
\sum_a J^a(\mu) + J^g(\mu)\=\frac12\,.
\label{eq:ji-sum-rule}
\ee
Combining Ji's sum rule \req{eq:ji-sum-rule} with the momentum sum rule of DIS
\be
\sum_a q_{20}^a(\mu) + g_{20}(\mu)\=1\,,
\label{eq:momentum-sum-rule}
\ee
one finds a sum rule for the second moments of $E$ \ci{teryaev99}
\be
\sum_a e_{20}^a(\mu) + e_{20}^g(\mu) \=0\,.
\label{eq:e-sum-rule}
\ee
These sum rules hold for any value of the renormalization scale.
One may also split the total angular momentum into its orbital and spin components. For a 
detailed discussion of gauge invariant definitions of the latter quantities it is referred to the 
review by Leader and Lorc\'{e} \ci{leader13}. The spin of a parton is related to the lowest 
moment of the GPD $\widetilde{H}$ at $\xi=t=0$. This moment reduces to the first moment of a 
polarized parton distribution, either for a quark or a gluon, 
\be
\Delta q^a_{10} \= \int_0^1 dx \Delta q^a(x)\,, \qquad \Delta g_{10}\=\int_0^1dx \Delta g(x)\,.
\label{eq:spin}
\ee
These moments are known from DIS. The spin of a flavor-a quark is $S^a=1/2 \Delta q_{10}^a$. In 
light-cone gauge, $\Delta g_{10}$ may be interpreted as the spin of the gluon \ci{ji96,leader13}.
Thus, one sees that the contribution of hard exclusive reactions to the 'spin puzzle' is the 
information on the GPD $E$. What is known on this GPD at present and a discussion of its properties 
is the topic of the following section.

\section{What do we know of the GPD $E$?}
\label{sec:E}
The zero-skewness GPD $E$ for a quark of flavor $a$ or the gluon may be parametrized as 
\ci{DK13,GK4,GK3} (for an overview see \ci{chia})
\be
E_{a}(x,\xi=0,t)\=e_{a}(x) \exp{[tg_{a}(x)]}\,,
\label{eq:zero-skewness-E}
\ee
assumed to be valid at the initial scale of $\mu_0=2\,\gev$. The profile function is parametrized as~\footnote
    {This parametrization of the profile function is supported by light-front holographic QCD \ci{teramond18};
     for a  similar parametrization see also \ci{moutarde18}.} 
\be
g_{a}(x)\=\big(B_a+\alpha'_a\ln{1/x}\big)(1-x)^3 + A_a x (1-x)^2
\label{eq:profile}
\ee
and the forward limit
\be
e_{a}(x)\=E_{a}(x,\xi=t=0)\= N_a \,x^{-\alpha_a}(1-x)^{\beta_a}(1+\gamma_a\sqrt{x})\,.
\label{eq:E-parametrization}
\ee
As is discussed in \ci{DK13,DFJK4} there is a strong $x - t$ correlation for GPDs of this type: 
At small $-t$ the GPD is large at small $x$ while, at large $-t$, the GPD is sharply peaked at a 
large value of $x$. Thus, the second term, proportional to $(1-x)^2$, is responsible for the
large $-t$ behavior of the moments of this GPD and ensures a finite distance between the active
parton and the cluster of spectators.

As is well-known the form factors of the nucleon are represented by the lowest moments of the
GPDs $H$ and $E$ for valence quarks. Thus, the Pauli form factor of the proton is given by
\be
F_2(t)\=\sum_{a_v} e_a F_2^a(t)
\label{eq:PauliFF1}
\ee
with the flavor-$a$ Pauli form factor 
\be
F_2^a(t)\=\int_0^1 dx E_{a_v}(x,\xi=0,t)
\label{eq:PauliFF2}
\ee
and the quark charges $e_a$ (in units of the positron charge).
In \ci{DK13,DFJK4} the relation \req{eq:PauliFF2} has been exploited to extract the GPD $E$ for 
valence quarks from the data on the nucleon form factors using the parametrization 
\req{eq:zero-skewness-E}. The normalization factor, $N_{a_v}$, of $E_{a_v}$ is fixed from 
\be
\int_0^1 dx e_{a_v}(x)\=\kappa_a
\ee
where $\kappa_a$ is the contribution of quark flavor $a$ to the anomalous magnetic moment of 
the proton ($\kappa_u\simeq 1.67$, $\kappa_d\simeq -2.03$). A possible contribution of strange 
quarks to the Pauli form factor is neglected. The parameters of $E$ for valence quarks 
appearing in Eqs.\ \req{eq:profile} and \req{eq:E-parametrization} can be found in \ci{DK13}. An 
analogous extraction of the GPD $H$ from the Dirac form factor has also been carried out in this 
work. The second moment of $E_{a_v}$ at $t=0$ is readily evaluated from \req{eq:e-moments}  and 
\req{eq:E-parametrization}. The following values have been obtained \ci{DK13}:
\be
e^{u_v}_{20}(\mu_0)\=0.163^{+0.018}_{-0.032}\,, \qquad 
                                          e^{d_v}_{20}(\mu_0)\=-0.122^{+0.028}_{-0.033}\,.
\label{eq:E-DK-1}
\ee
It is to be stressed that the values of $e_{20}^{a_v}$ depend crucially on the parameters 
$\beta_a$ and $\gamma_a$.  The sum of the second  moments
\be
e^v_{20}(\mu)\=e^{u_v}_{20}(\mu) + e^{d_v}_{20}(\mu)
\label{eq:ev}
\ee
is
\be
e^{v}_{20}(\mu_0)\=0.041^{+0.011}_{-0.053}\,.
\label{eq:E-DK-2}
\ee
 
Further information on $E$ is obtained from measurements of deeply virtual exclusive vector meson 
(DVVP) and photon (DVCS) electroproduction with a transversely polarized target. To study such
processes within the handbag approach the skewness dependence of the GPDs is needed. In
\ci{GK3,GK4,GK7} the zero-skewness GPD \req{eq:zero-skewness-E} is used as input to the double 
distribution ansatz \ci{musatov00} with which the skewness dependence of $E$ is generated~\footnote{
     In \ci{DK13,GK3,GK4,GK7} all zero-skewness GPDs are parametrized as in \req{eq:zero-skewness-E},
     \req{eq:E-parametrization} and \req{eq:Regge-profile}. For all of them the skewness dependence 
     is generated with the help of the double distribution ansatz. In some cases, as for instance for 
     the GPD $H$, the forward limits reduce to parton densities.}.  
For hard exclusive processes for which $-t$ is typically rather small it suffices to use the 
so-called Regge-like profile function 
\be
g_{Ra}(x)\=B_a+\alpha'_a\ln{1/x}
\label{eq:Regge-profile}
\ee
instead of \req{eq:profile}. Detailed investigation \ci{GK7} revealed that the $\sin{(\phi-\phi_s)}$ 
modulation of the transverse target spin asymmetry, $A_{UT}$, is most sensitive to the GPD $E$. Here, 
$\phi$ is the azimuthal angle between the lepton and the production plane while $\phi_s$ specifies 
the orientation of the target spin vector with respect to the lepton plane. The spin vector is
perpendicular to the direction of the virtual photon.
This modulation of $A_{UT}$ is under control of the imaginary part of the interference between proton
helicity flip and non-flip amplitudes 
\be
A_{UT}^{\sin{(\phi-\phi_s)}}\frac{d\sigma}{dt}\=
                                  -2\, {\rm Im}\,\sum_\nu M^*_{\nu -,\nu +}M_{\nu +,\nu +}
\label{eq:AUT}
\ee
where $\nu$ marks the (longitudinal or transverse) polarization of the virtual photon and the 
vector meson (or real photon). Explicit helicities are labeled by their signs only. In the handbag 
approach each of the amplitudes in \req{eq:AUT} is expressed by a convolution,
${\cal K}_{j\nu}$, of a suitable flavor combination of the relevant GPD, $K^j$, and a perturbatively
calculable amplitude, $H^j_{\nu\lambda,\nu\lambda}$, for the hard subprocess
$\gamma^*(\nu) q(\lambda) \to P_j(\nu) q(\lambda)$ ($\nu$ and $\lambda$ denote the helicity
of the respective particle $P_j$, $j=V,\gamma$)
\be
   {\cal K}_{j\nu}\=\sum_\lambda \int_{-1}^1 dx H^j_{\nu\lambda,\nu\lambda} K^j
 \ee
 The interference between the proton helicity flip and non-flip amplitudes in \req{eq:AUT} is
 therefore proportional to the interference of the convolutions of the GPDs $E$ and $H$
\be
{\rm Im}\,\big[M^*_{\nu -,\nu +}M_{\nu +,\nu +}\big] \sim \frac{\sqrt{t-t_0}}{2m} 
                                      {\rm Im}\big[{\cal E}^*_{j\nu}{\cal H}_{j\nu}\big]\,.
\label{eq:interference}
\ee   

In electroproduction forward scattering occurs at $t=t_0$ where $t_0$ is related to the skewness by 
\be
t_0\=-4m^2\frac{\xi^2}{1-\xi^2}
\ee
($m$ being the proton mass). Using the GPD $H$ for quarks and gluons as determined in \ci{DK13,GK3} 
which dominates the differential cross section, $d\sigma/dt$, for DVVP and DVCS, 
and the above discussed $E$ for valence quarks in the evaluation of the asymmetry \req{eq:AUT} for
$\rho^0$ production, one obtains reasonable agreement with the HERMES \ci{hermes-rho} (at $W=5\,\gev$ 
and a photon virtuality, $Q^2$, of $2\,\gev^2$) and COMPASS \ci{compass-rho} (at $W=8.1\,\gev$, 
$Q^2\=2.2\,\gev^2$) data (see Fig.\ 5 in \ci{GK7})~\footnote{
         In electroproduction the photon virtuality is identified
         with the renormalization scale $\mu^2$.}.
In this calculation the values $B_{u_v}=B_{d_v}=0$ and $\alpha'_{u_v}=\alpha'_{d_v}=0.9\,\gev^{-2}$ 
are taken in \req{eq:Regge-profile} \ci{GK3,GK4}. Thus, it seems that $E$ for sea quarks and gluons 
does not noticeable contribute to the asymmetry \req{eq:AUT} for $\rho^0$ production. This can be 
understood: Since $e^v_{20}$ is rather small (see \req{eq:E-DK-2}) for $\mu\simeq \mu_0$ the sum rule 
\req{eq:e-sum-rule} tells us that 
\be
e_{20}^g(\mu_0) \simeq - 2\big[e_{20}^{\bar{u}}(\mu_0) + e_{20}^{\bar{d}}(\mu_0)
                 + e_{20}^{\bar{s}}(\mu_0)\big] \=-e_{20}^{sea}(\mu_0)\,.
\ee 
For parametrizations of $E(x,\xi=t=0)$ without nodes the approximate equality of the second 
moments for gluons and sea quarks holds for other moments and even for the convolutions with a 
hard subprocess amplitude too. Consequently, the sea quark and gluon contributions to ${\cal E}$ 
cancel to a large extent. From this consideration it is also evident that in hard exclusive $\phi$ 
production to which only gluons and strange quarks contribute, a small $\sin{(\phi-\phi_s)}$ 
modulation of $A_{UT}$ is to be expected. 

Since gluons do not contribute to DVCS at leading order (LO) the partial cancellation of the gluon 
and sea-quark contributions from $E$ does not occur and, hence, $E_{sea}$ may become visible in $A_{UT}$
for DVCS. Indeed the valence quark contribution from $E$ to the $\sin{(\phi-\phi_s)}$ modulation
of $A_{UT}$ does not match well the data measured by the HERMES collaboration \ci{hermes-aut-dvcs} 
(see Fig.\ 8 in \ci{KMS}). An additional negative contribution from $E_{sea}$ seems to be required by 
the data although the experimental errors are large. The possibility of a negative $E_{sea}$ is also 
supported by the HERMES data on the $sin{(\phi-\phi_s)}\cos{\phi}$ modulation \ci{hermes-aut-dvcs} 
which is under control of the interference between the Bethe-Heitler and the DVCS amplitudes. 
In order to proceed one has to assume a flavor-symmetric light-quark sea in $E$; the available 
information on $E$ does not allow for a more detailed parametrization at present. On this assumption 
a restriction of the magnitude of $E_{sea}$ follows from a positive bound for $E_s$ as has been 
advocated for by Diehl and Kugler \ci{diehl-kugler}. The bound, derived by Burkardt \ci{burkhardt}, 
reads
\be
\frac{b^2}{m^2}\left(\frac{\partial \hat{e}_s(x,b)}{\partial b^2}\right)^2\leq \hat{s}^{2}(x,b)
                - \Delta \hat{s}^2(x,b)
\label{eq:bound-1}
\ee
where $\hat{e}_s$, $\hat{s}$ and $\Delta \hat{s}$ are Fourier conjugated to the zero-skewness 
strange-quark GPDs $E_s, H_s$ and $\tilde{H}_s$, respectively. The impact parameter, ${\bf b}$, 
is canonically conjugated to the two-dimensional momentum transfer ${\bf \Delta}$ ($\Delta^2=t$). 
This bound ensures positive semi-definite densities of partons in the transverse plane. 
For $\widetilde{H}_s$ only the forward limit, the polarized strange-quark density, $\Delta s(x)$, 
is known, not its $t$ dependence. However, since $\Delta s(x)$ is quite small \ci{DSSV09} it seems 
reasonable to ignore the polarized strange-quark contribution to the bound \req{eq:bound-1}. Doing 
so the bound is slightly weakened. According to \ci{DK13,GK4,GK3} the GPD $H_s$ is
parametrized analogously to $E$, see \req{eq:zero-skewness-E}--\req{eq:E-parametrization} with
the forward limit, $\xi, t\to 0$, taken as the strange-quark parton density. With the help of
these parametrizations of $H_s$ and $E_s$, one can put
the simplified bound \req{eq:bound-1} into a more convenient form. As 
shown in \ci{DFJK4} the bound is most stringent for $b^2=2g_sf_s/(f_s-g_s)$ (denoting the profile 
function of $H_s$ by $f_s$). For this value of $b$ the bound reads 
\be
e_s^2(x) \leq 8m^2 e \left(\frac{g_s(x)}{f_s(x)}\right)^3 \big(f_s(x)-g_s(x)\big)\, s(x)^2\,.
\label{eq:bound-2}
\ee
Immediate consequences of this bound are
\be
g_s(x)<f_s(x)\,, \qquad   \beta_s\geq 1+\beta_{h_s}
\label{eq:consequences}
\ee
where $\beta_s$ is defined in \req{eq:E-parametrization} and $\beta_{h_s}$ is the power
of $1-x$ with which the strange-quark parton density, $s(x)$, tends to zero for $x\to 1$. 
As is \ci{GK3} and in agreement with the PDFs given in \ci{ABM11}, the power $\beta_{h_s}$ is taken to be 5.
In order to respect the r.h.s. of \req{eq:consequences} $\beta_s=7$ is therefore chosen. It has been shown
on the basis of perturbative QCD arguments \ci{yuan04} that $E$ for valence quarks behaves as $(1-x)^5$
for $x\to 1$. Sea quarks are expected to fall faster. Thus, the chosen value of $\beta_s$ appears to be plausible.

In order to exploit the bound \req{eq:bound-2} the parameters of $E_s$ are to be specified whereas
$H_s$ can be taken from \ci{GK3}. The low-$x$ behavior of $E$ for strange quarks and gluons 
is under control of the gluon, i.e.\ a Pomeron-like, Regge trajectory as is the case for $H_s$ and $H_g$. 
According to Regge phenomenology the trajectory should be the same for $E$ and $H$ but not the residues. 
The corresponding parameters can directly be read off from HERA data on $\rho^0$ and $\phi$
electroproduction \ci{h1,zeus}
\be
\alpha_s\=1.1\,, \qquad \alpha'_s\=0.15\,\gev^{-2}\,.
\ee 
at the scale $\mu_0$. The slope parameter $B_{h_s}$ is taken to be $2.15\,\gev^{-2}$ \ci{GK3}. The restriction 
$g_s<f_s$, see \req{eq:consequences}, requires the parameter $A$ in the profile function \req{eq:profile} 
of $E_s$ to be larger or equal to that of $H_s$. Taking equal values for this parameter as well as 
$B_s=0.9 B_{h_s}$ \ci{diehl-kugler} and $\gamma_s=0$, the bound \req{eq:bound-2} is saturated for 
$N_s=\pm 0.138$ ( at $x\simeq 0.12$). The Regge-like profile \req{eq:Regge-profile} leads to a similar
result. With the above value of the normalization factor, $N_s$, the second moment of $E$ for strange 
quarks at $t=0$ is
\be 
|e_{20}^s (\mu_0)|\leq 0.024\,. 
\ee
The sign of $N_s$ is not determined by the bound \req{eq:bound-2} in which the square of $e_s(x)$ appears, 
but one can take into account that the DVCS data on $A_{UT}$ favor a negative $E^{sea}$ as is discussed 
above. On the assumption of a flavor symmetric light-quark sea one finds the following restriction of
$e_{20}^{sea}$:
\be
e_{20}^{sea}(\mu_0)\=6\, e_{20}^s(\mu_0)\= -0.073 \pm 0.073\,,
\label{eq:S-0}
\ee
The uncertainty of  $e_{20}^{sea}$ is not a statistical error. It rather defines the upper and lower 
limits of $e_{20}^{sea}$; any value between them is possible at present (for each choice $e_{20}^g$ is fixed
by the sum rule \req{eq:e-sum-rule}). The full flavor-singlet combination is restricted by
\be
\Sigma_e(\mu_0)\=-e_{20}^g(\mu_0)\=\sum_a e_{20}^a(\mu_0)\=-0.032^{+0.084}_{-0.126}\,.
\label{eq:sigma-0}
\ee

From \req{eq:E-DK-2} and \req{eq:S-0} one sees that at the scale $\mu_0$ 
\be
e_{20}^u+e_{20}^d+e_{20}^{\bar{u}}+e_{20}^{\bar{d}}\=e^v_{20}+\frac23\,e_{20}^{sea}
\label{eq:u+d}
\ee
is compatible with zero within large uncertainties. This result is in accordance 
with the large $N_c$ result that $E_u+E_d$ is suppressed by $1/N_C$ \ci{goeke}.

Admittedly the present uncertainty of $e_{20}^{sea}$ is large as a consequence of the 
large experimental errors of $A_{UT}^{\sin{(\phi-\phi_s)}}$ for DVCS. A more accurate measurement 
of this observable would immediately reduce the uncertainty of $e_{20}^{sea}$. 

Finally, a few comments concerning $E_g$ are in order. The normalization, $N_g$, of $E_g$ 
\req{eq:zero-skewness-E} is fixed via the sum rule \req{eq:e-sum-rule} \ci{GK4}. The bounds on $E_g$, 
analogous to \req{eq:bound-1} and \req{eq:bound-2}, are far from being saturated. Independent 
information on $E_g$ would also be of interest. Such information may be obtained from a measurement 
of the transverse target polarization, $A_N$, in $J/\Psi$ photoproduction \ci{metz} which plays the 
role of $A_{UT}^{\sin{(\phi-\phi_s)}}$ in electroproduction. Since $H_g$ and $E_g$ provide the 
dominant contribution to $J/\Psi$ photoproduction $A_N$ can be written as   
\be
A_N \sim \frac{\sqrt{t-t_0}}{2m}\, \frac{|{\cal E}_{J/\Psi}|}{|{\cal H}_{J/\Psi}|}\,\sin{\delta}
\label{eq:AN}
\ee   
where $\delta$ is the relative phase between the two convolutions. Since the power, $\alpha_g$, 
of $x$ is expected to be the same in the parametrization \req{eq:E-parametrization} and in the 
analogous one for $H_g$, the relative phase $\delta$ is likely small except perhaps at small energies 
provided that the $x$-dependencies of $H_g$ and $E_g$, apart from the leading power of $x$, are very 
different, e.g.\ if $e_g(x)$ has nodes. In this case, as is shown in \ci{metz}, values of $A_N$ as large as 
$\simeq 0.1$ are possible. A precise measurement of $A_N$ in $J/\Psi$ photoproduction may therefore 
provide useful information on $E_g$. Another interesting observable is the spin correlation between 
a longitudinally polarized beam and sideways polarization of the recoil proton, $A_{LS}$. It is 
related to convolutions as in \req{eq:AN} but being proportional to $\cos{\delta}$.

\section{Evolution of the second moments of $E$}
\label{sec:evolution}
Since the evolution of the forward limits of the GPDs $H$ and $E$ are controlled by the same  
anomalous dimensions the corresponding moments fulfill the same differential equation. Thus, 
the flavor-singlet 
combination of $E$ mix with the gluon GPD, $E_g$, under evolution. To next-to-leading order (NLO), 
the differential equation for their second moments, $\Sigma_e=\sum_a e_{20}^a$ and $e_{20}^g$, reads
\ci{buras80}
\be
\frac{d}{dt} {\Sigma_e \choose e^g_{20}} \= - \beta_0\frac{\alpha_s}{4\pi}\Big[{\cal D}^{(0)}
                   + \frac{\alpha_s}{4\pi} {\cal D}^{(1)}\Big]{\Sigma_e \choose e_{20}^g}
\label{eq:S-diff-eq}
\ee
where $t=\ln{(\mu^2/\LQCD^2)}$. The NLO effective coupling constant is given by
\be
\frac{\alpha_s}{4\pi}\=\frac1{\beta_0t}\,\Big(1-\frac{\beta_1}{\beta_0}\,
                                \frac{\ln{t}}{t}\Big)\,.
\ee
The renormalization group functions are
\be
\beta_0\=11-\frac23 n_f\,,  \qquad \beta_1\=102-\frac{38}{3}n_f
\ee
where $n_f$ denotes the number of active flavors. The LO and NLO matrices of the anomalous 
dimensions, ${\cal D}^{(i)}$, are quoted in \ci{buras80,floratos79} (see also \ci{haegler10}). 
They possess the remarkable property~\footnote
       {The matrix elements $d_{ab}^{(i)}$ are the usual anomalous dimensions
        $\gamma_{ab}^{(i)}$ divided by $\beta_0$.}
($i=0,1$)
\be
d_{gq}^{(i)}\=-d_{qq}^{(i)}\,, \qquad d_{qg}^{(i)}\=-d_{gg}^{(i)}\,,
\label{eq:ano-dim-relations}
\ee
which guarantees that the sum rules \req{eq:ji-sum-rule}, \req{eq:momentum-sum-rule} and 
\req{eq:e-sum-rule} hold at all scales. As a consequence of the relations in \req{eq:ano-dim-relations} 
one of the eigenvalues of the anomalous dimension matrices is zero while the other one is
\be
d_S^{(i)}\= d_{qq}^{(i)} + d_{gg}^{(i)}\,.
\label{eq:singlet-ano-dim}
\ee
Because of the sum rule \req{eq:e-sum-rule} implying
\be
e_{20}^g(\mu)\=-\Sigma_e(\mu)\,,
\ee
and of the property \req{eq:ano-dim-relations} the matrix equation \req{eq:S-diff-eq} formally falls 
into two identical scalar ones which have the same structure as the differential equation for the 
second moment of $E$ in the flavor non-singlet case. Hence,
\be
\frac{d}{dt} M_{j}\=- \beta_0\frac{\alpha_s}{4\pi}\Big[d_j^{(0)} 
                          + \frac{\alpha_s}{4\pi}d_j^{(1)}\Big]
\label{eq:diff-eq}
\ee
(for $j=S, NS$; $M_{S}=\Sigma_e$, $M_{NS}=e^v_{20}$) where $d_{NS}^{(i)}=d_{qq}^{(i)}$ and $d_{S}^{(i)}$ is 
defined in \req{eq:singlet-ano-dim}. The numerical values of the anomalous dimensions, evaluated to NLO 
in the $\overline{MS}$ scheme and for $n_f=4$, are \ci{haegler10,buras80}
\be
d_{NS}^{(0)}\=\frac{32}{75}\,, \quad d_{S}^{(0)}\=\frac{56}{75}\,, \quad
d_{NS}^{(1)}\=4.286\,, \quad d_{S}^{(1)}\=6.955\,.
\label{eq:NLO-parameter}
\ee
Since the moments of $E$ will be evolved up to $100\,\gev^2$, the charm treshold is passed.
  Therefore four flavors are used in the evolution equation.

The solution of \req{eq:diff-eq} is
\be
M_{j}(\mu)\= R_j(\mu)\,M_{j}(\mu_0)\,e^{-s d_j^{(0)}} 
\ee
where
\be
R_j(\mu)\=1-\frac{\beta_1}{\beta_0^2}\,d_j^{(0)}
                    \Big[\frac{\ln{t}}{t}-\frac{\ln{t_0}}{t_0}\Big] 
          - \Big(\frac{\beta_1}{\beta_0^2}\,d_j^{(0)} - \frac{d_j^{(1)}}{\beta_0}\Big)\,
            \Big[\frac1{t}-\frac1{t_0}\Big]
\label{eq:R}
\ee
up to corrections of order $1/t^2$. The variable $s$ is defined as 
\be
s=\ln{(t/t_0)}\,. 
\ee 

The charm contribution at the initial scale is assumed to be zero
\be
e_{20}^c(\mu_0)\=0\,.
\ee
The charm quark is assumed to be massless as usual in this type of analysis. This assumption likely
leads to an overestimate of the strength of the charm density. It is convenient to introduce
the following flavor-non-singlet combination \ci{buras80}
\be
 \Delta_{sc}(\mu)\= e_{20}^s(\mu)-e_{20}^c(\mu) \= 
              \frac16 e_{20}^{sea}(\mu_0)\,R_{NS}(\mu)\,e^{-s d_{NS}^{(0)}}\,.
\label{eq:result-sc}
\ee
In detail~\footnote{ 
    The LO part of this analysis is analogous to the parton-densities analysis performed by Buras 
    and Gaemers \ci{buras-gaemers}. The density analysis is equivalent to the analysis of the second moments
   of $H(x,\xi=t=0)$.},
the flavor-non-singlet combination $e^v_{20}$, defined in \req{eq:ev}, evolves as 
\be
e^v_{20}(\mu) \= e^v_{20}(\mu_0)\,R_{NS}(\mu)\,e^{-s d_{NS}^{(0)}}
\label{eq:result-V}
\ee
and  flavor-singlet combination as
\be
\Sigma_e(\mu)\=e^v_{20}(\mu) + e_{20}^{sea}(\mu) + 2e_{20}^c(\mu)\=-e_{20}^g(\mu) \=
                                          \Sigma_e(\mu_0)\,R_S(\mu)\,e^{-s d_S^{(0)}}\,.
\label{eq:results-S}
\ee  
Combinations of these three relations allow to disentangle $e_{20}^{sea}$ and $e_{20}^c$:
\ba
e_{20}^{sea}(\mu)&=&\frac34\Sigma_e(\mu_0)\,R_S(\mu)\,e^{-s d_S^{(0)}}
            - \frac14\big[3e^v_{20}(\mu_0)-e_{20}^{sea}(\mu_0)\big]\,R_{NS}(\mu)\,e^{-s d_{NS}^{(0)}} \nn\\
e_{20}^c(\mu)&=&\frac18\Sigma_e(\mu_0)\,\big[R_S(\mu)\,e^{-s d_S^{(0)}}
                              - R_{NS}(\mu)\,e^{-s d_{NS}^{(0)}}\big]\,.
\label{eq:results-S-C}
\ea
For the scales of interest here the difference between NLO and LO evolution is small, of the order 
of a few per cent. As shown in \ci{haegler10} for scales smaller than $1\,\gev$ the differences are 
substantial. The NLO factors $R_j$, defined in \req{eq:R}, are of marginal importance; they deviate 
less than $2\%$ from unity for scales of interest.

The values of $e_{20}$ at the initial scale,  $\mu_0$, given in Eqs.\ \req{eq:E-DK-1}, 
\req{eq:E-DK-2}, \req{eq:S-0} and \req{eq:sigma-0}, can now be evolved to other scales. The 
results, obtained with $n_f=4$ and $\LQCD=270\,\mev$, are shown in Figs.\ \ref{fig:valence} and 
\ref{fig:gluon}~\footnote{
    With this value of $\LQCD$ the evolution of the second moments of $H(x,\xi=t=0)$ leads to
    the same results as obtained from integrating the evolved ABM11 PDFs \ci{ABM11} obtained from 
    the Durham PDF plotter.}. 

\begin{figure}[t]
\centering
\includegraphics[width=0.48\tw]{fig-2nd-moments-valence-quark.epsi}
\includegraphics[width=0.48\tw]{fig-2nd-moments-gluon.epsi}
\caption{\label{fig:valence} Left: The second moments of the valence-quark GPDs $H$ and $E$ at 
$\xi=t=0$ versus the scale. The bands indicate the uncertainties of the $E$ moments.}
\caption{\label{fig:gluon} Right: The second moments of the GPD $E$ for gluons and light sea quarks 
at $\xi=t=0$. The upper (lower) and lower (upper) edge of the gluon (sea-quark) band is 
evaluated from $e_{20}^{sea}(\mu_0)=-0.146$ and  $0$, respectively. The thick solid and dashed lines from 
$e_{20}^{sea}(\mu_0)=-0.073$. The bands indicate the uncertainties of the gluon and sea-quark 
moments.}
\end{figure}

As a consequence of the detailed, precise information about $E$ for valence quarks \ci{DK13} 
its second moments have rather small errors. On the other hand, due to the large experimental
errors of the $\sin{(\phi-\phi_s)}$ modulation of $A_{UT}$ in DVCS \ci{hermes-aut-dvcs} the 
uncertainty of $e_{20}^{sea}$ and, hence, that of $e_{20}^g$ are large and, forced by the bound
\req{eq:bound-2}, these moments are small in absolute value ($e_{20}^g<<h_{20}^g$). Since the 
sum rule \req{eq:e-sum-rule} is used to fix $e_{20}^g$ its uncertainty is strongly correlated with that 
of $e_{20}^{sea}$. As is to be seen from Figs.\ \ref{fig:valence} and \ref{fig:gluon} and as is 
evident from Eqs.\ \req{eq:result-sc} - \req{eq:results-S-C}, all second moments of $E$ tend to 
zero for asymptotically large scales. This is in accordance with the behavior of $E$ for 
asymptotically large scales \ci{diehl03,goeke}.

\section{Parton angular momenta}
\label{sec:J}
In order to evaluate the parton angular momenta \req{eq:ji} one has to add the second
moments of $E$ to the information from DIS, namely the second moments of the parton densities 
\req{eq:density-moments}. In Tab.\ \ref{tab:density-moments} these moments, evaluated from
various sets of densities \ci{ABM11}-\ci{NNPDF}, are compiled. Since the GPD analysis of
the nucleon form factors \ci{DK13} is based on the ABM11 densities \ci{ABM11} (NLO, $n_f=4$, 
$\overline{MS}$ scheme), they have also to be applied in the evaluation of the angular 
momenta for consistency. The considerable spread of the values quoted in Tab.\ \ref{tab:density-moments} 
is regarded as an estimate of the uncertainties of the ABM11 values.  
\begin{table*}[t]
\renewcommand{\arraystretch}{1.4} 
\begin{center}
\begin{tabular}{| c || c | c | c | c | c | c | c |}
\hline 
PDF  & $u_v$  & $d_v$ & $\bar{u}$ &  $\bar{d}$ & $s$ & $\Sigma_q$ & g \\[0.2em]
\hline 
ABM \ci{ABM11}& 0.297 & 0.115 & 0.031 & 0.039 & 0.017 & 0.586 & 0.409  \\[0.2em]
CT \ci{CT10}  & 0.287 & 0.118 & 0.029 & 0.036 & 0.020 & 0.575 & 0.421 \\[0.2em]
GJR \ci{GJR}  & 0.280 & 0.116 & 0.032 & 0.040 & 0.011 & 0.562 & 0.447\\[0.2em]
HERAPDF \ci{HERA}& 0.284& 0.105& 0.037& 0.046 & 0.022 & 0.599 & 0.373 \\[0.2em]
MSTW \ci{MSTW}& 0.282 & 0.115 & 0.032 & 0.038 & 0.017 & 0.571 & 0.424 \\[0.2em]
NNPDF \ci{NNPDF}&0.290& 0.124 & 0.030 & 0.037 & 0.010 & 0.568 & 0.429 \\[0.2em]
\hline
\end{tabular}
\end{center}
\caption{The second moments of the parton densities at the scale $\mu_0$ evaluated from various 
sets of PDFs. The charm contribution is not shown. The deviation of the sum of 
$\sum_a q^a_{20}$ and $g_{20}$ from 1 is less than $1\%$. The table is prepared with the 
help of table 17 given in \ci{DK13}.}
\label{tab:density-moments}
\renewcommand{\arraystretch}{1.0}   
\end{table*} 

\begin{figure}[t]
\centering
\includegraphics[width=0.6\tw]{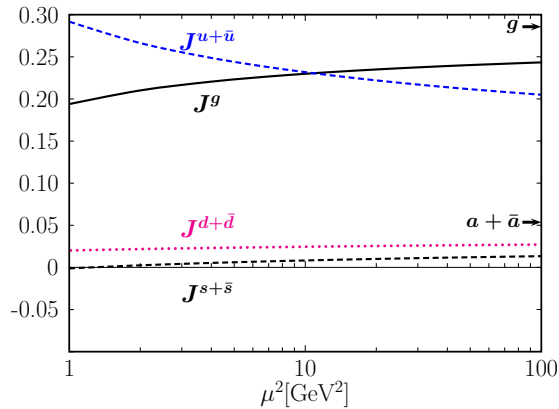}
\caption{Evolution of the parton angular momenta for quarks and gluons ($e^{sea}_{20}(\mu_0)=-0.073$). 
The charm contribution is not shown. The arrows indicate the asymptotic values of $J$ for gluons and 
for a single quark flavor.}
\label{fig:J}
\end{figure}

The results for the angular momenta at the scale $\mu_0$ are:
\ba
J^{u+\bar{u}} &=& 0.249^{+0.022}_{-0.036} \nn\\
J^{d+\bar{d}} &=& 0.024^{+0.033}_{-0.033} \nn\\
J^{s+\bar{s}} &=& 0.005^{+0.014}_{-0.014} \nn\\
J^g           &=& 0.221^{-0.067}_{+0.084}
\label{eq:J}
\ea  
As usual the sum of the angular momentum a quark and its antiquark carries, is quoted here. 
In order to avoid confusion this sum is denoted by $J^{a+\bar{a}}$. This is to be contrasted with 
the notation $J^a$ for the sum to be frequently found in the literature. There is also a little 
contribution to the proton spin from charm quarks. The uncertainties of the quark contributions are not 
statistical but rather define upper and lower bounds as is the case for the second moments of $E$. 
It is to be stressed that the angular momentum carried by the gluons is, in the end, obtained from 
the sum rule \req{eq:ji-sum-rule}. The evolution of the face values of the angular momenta \req{eq:J} are 
displayed in Fig.\ \ref{fig:J}. Except for the case of the gluon the contributions from $E$ 
are substantial. For $J^{u+\bar{u}}$ the contributions from $H$ and $E$ add while for $d$ and 
$s$ quarks they have opposite signs and cancel each other to a large extent. This feature makes it 
clear that $J^{u+\bar{u}}$ and $J^g$ are large while $J^{d+\bar{d}}$ and $J^{s+\bar{s}}$ are small.  Thus,
gluons and $u$-quarks share almost equally the spin of the proton for intermediate scales around 
$10\,\gev^2$. For lower scales $J^{u+\bar{u}}$ is larger than $J^g$, for larger scales smaller.   

Ji \ci{ji96,ji95} has shown that the asymptotic limits of the angular momenta are  
\ba
J^{a+\bar{a}}&\stackrel{\mu^2\to\infty}{\longrightarrow}& \frac12\frac{d_{gg}^{(0)}}{d_S^{(0)}}
                            \=\frac12 \frac{3}{16+3n_f}  \nn\\
J^g &\stackrel{\mu^2\to\infty}{\longrightarrow}& \frac12\frac{d_{qq}^{(0)}}{d_S^{(0)}}
                    \=\frac12 \frac{16}{16+3n_f}\,.
\label{eq:Jasymp}
\ea
The limiting values are approached logarithmically. They solely come from the GPD $H$. In fact, aside 
of the prefactor $1/2$, the asymptotic values of $J$ are those of the second moments of $H$. The latter 
tell us the asymptotic partition of the proton momentum \ci{gross}. For the scales displayed 
in Fig.\  \ref{fig:J} the angular momenta of the gluon, down and strange quarks are 
somewhat below their asymptotic values whereas $J^{u+\bar{u}}$ is substantially larger than the value
quoted in \req{eq:Jasymp}.

\begin{figure}[th]
\centering
\includegraphics[width=0.6\tw]{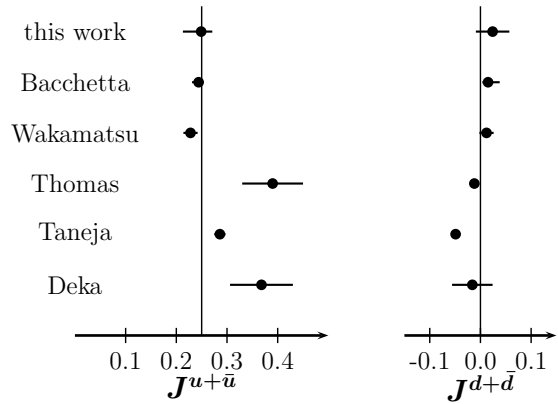}
\caption{Various results on $J^{u+\bar{u}}$ and $J^{d+\bar{d}}$ at the scale $\mu_0$. The results are 
taken from Bacchetta \ci{bacchetta}, Wakamatsu \ci{wakamatsu}, Thomas \ci{thomas}, 
Taneja \ci{liuti} and Deka \ci{deka}.}
\label{fig:J-comparison}
\end{figure}

In Fig.\ \ref{fig:J-comparison} the results \req{eq:J} on $J$ for $u$ and $d$ quarks are compared 
to other ones. Good agreement is to be observed with the values quoted in \ci{bacchetta,wakamatsu}.
These analyses are similar to the one presented in this work in so far as the angular momenta
are calculated from the PDFs and estimates of the GPD $E$. In \ci{bacchetta} a model-dependent
relation between  $E$ and the Sivers function, extracted from polarized semi-inclusive DIS, 
is used while in \ci{wakamatsu} the difference of $e_{20}^{u_v}$ and $e_{20}^{d_v}$ is taken from lattice 
QCD and the sum, $e_{20}^v$, is assumed to be bounded by zero and by the sum of the corresponding first 
moments of $E$ ($\kappa_p+\kappa_n=(\kappa_u+\kappa_d)/3)=-0.12$) which hardly overlaps with the range
specified by \req{eq:E-DK-2}. Other results \ci{thomas,liuti,deka} for $J^{u+\bar{u}}$ and $J^{d+\bar{d}}$ 
differ markedly 
from \req{eq:J}. In particular the large values of $J^{u+\bar{u}}$ in these paper suggest small values 
for the angular momentum of the gluon, Indeed, the lattice QCD calculation of Deka \ci{deka} provides 
$J^g=0.139\pm 0.038$, clearly smaller than the value of $J^g$ quoted in \req{eq:J}. On the other hand,
a QCD sum rule estimate leads to a larger value: $J^g\simeq 0.27$ \ci{balitsky}. 

For comparison the angular momenta for valence quarks are also quoted \ci{DK13}:
\be
J^{u_v}\=0.230^{+0.009}_{-0.024}\,, \qquad J^{d_v}\=-0.004^{+0.010}_{-0.016}
\ee
at the scale $\mu_0$. These values are in agreement with recent lattice QCD results \ci{lhpc,qcdsf07}
but differ from those quoted in \ci{deka}.

There are experimental results on the parton angular momenta extracted from DVCS data \ci{hallA-mazouz,
hermes-ye} which, within very large errors, are in agreement with the above quoted results.
These results are however strongly model-dependent. They rely for instance on the assumption
of a proportionality between $e_a(x)$ and $q_a(x)$ \ci{ellinghaus} which is in conflict with the 
behavior of the Pauli and Dirac nucleon form factors \ci{DK13} and with perturbative QCD arguments 
\ci{yuan04}.  

\section{Summary and discussion}
\label{sec:summary}
The information about the GPD $E$ which, on the basis of a parametrization, can be
extracted from current experimental data on the nucleon form factors and on polarized hard exclusive
reactions, is collected. Using in addition to this information a positivity bound for the Fourier
transform of $E$ for strange quarks and assuming a flavor symmetric sea in $E$, one can constrain
$E$ for all quarks although with occasionally large uncertainties. 
From this information the second moments of $E$ for quarks are evaluated and, completed by $e^g_{20}$
obtained from the sum rule \req{eq:e-sum-rule}, their scale dependence is studied. Combining these  
results with the second moments of the parton densities, $q_{20}$, the angular momenta carried by the partons
inside the proton, are evaluated.
It turns out that for scales of about $10\,\gev^2$ $u+\bar{u}$ and the gluon 
share almost equally the proton spin while the other flavors do not contribute much to it.    
At lower scales $J^{u+\bar{u}}$ is larger than $J^g$, at larger scales smaller.
A reduction of the uncertainties of $E$ is necessary in order to obtain more precise
information about the parton angular momenta. Data on the $\sin{(\phi-\phi_s)}$
and/or $\sin{(\phi-\phi_s)} \cos{\phi}$ modulations of the transverse target spin asymmetry 
in DVCS more accurate than the HERMES ones \ci{hermes-aut-dvcs} would  allow this.
Such measurements can be performed at the upgraded Jefferson Lab. 

The lowest moments of the polarized parton distributions, as for instance those derived 
in \ci{DSSV09}, represent the spin of the parton \req{eq:spin}: $S^a=\Delta q_{10}^a/2$ and
$S^g=\Delta g_{10}$ \ci{ji96,leader13}. Subtracting these contributions from the angular momenta, 
one obtains the orbital angular momenta of the partons. Since the main interest of this paper 
is focussed on the discussion of the GPD $E$ the extraction of the parton spin and orbital angular 
momenta is beyond the scope of this article. However, it may be of interest to quote some
important features: 
In a recent analysis \ci{DSSV14} which includes data on the double-spin asymmetry, $A_{LL}$, in
jet \ci{star} and $\pi^0$ production \ci{phenix} at RHIC a positive value for the gluon spin
has been obtained: $\Delta g_{10}(\mu_0)\simeq 0.1$, subject to a substantial uncertainty. This result
on $\Delta g_{10}$ implies a large difference $J^g-\Delta g_{10}\simeq 0.12$ 
which, in light-cone gauge, may be regarded as the gluon orbital angular momentum \ci{ji96,leader13}. 
The orbital angular momenta of $u$ and $d$ valence quarks are
\be
L^{u_v}\=-0.141^{+0.025}_{-0.033}\,, \qquad L^{d_v}\=0.113 \pm 0.025
\ee
and those of $u+\bar{u}$ and $d+\bar{d}$ 
\be
L^{u+\bar{u}}=-0.158^{+0.025}_{-0.033}\,, \qquad  L^{d+\bar{d}}=0.252\pm 0.025
\ee
at the scale $\mu_0$. These results have opposite signs to those obtained at hadronic scales 
\ci{thomas,myhrer07}. This difference is one of the key issues of the so-called 'spin crisis'. 
It is a consequence of the scale dependence of $J$, $L$ and the parton spin as it is understood 
now \ci{thomas,haegler10}: NLO (and NNLO) evolution indicates a change of sign of $L^{u+\bar{u}}$ 
and $L^{d+\bar{d}}$ at a low scale of about $0.3\,\gev^2$. Another interesting fact are the small 
values of $J^{d_v}$ and  $J^{d+\bar{d}}$ which are generated by an almost perfect cancellation of 
the orbital angular momentum and the spin of the $d$ quark. 

{\it Acknowledgement:} It is a pleasure to thank Markus Diehl for valuable comments.
 
\end{document}